\documentclass[pdflatex,sn-mathphys]{sn-jnl}

\usepackage{lineno,hyperref}
\modulolinenumbers[5]
\usepackage[utf8]{inputenc}
\usepackage[T1]{fontenc}
\usepackage{graphicx}
\usepackage{dcolumn}
\usepackage{bm}
\usepackage{epsfig}
\usepackage{subfigure}
\usepackage{graphics}
\usepackage{epstopdf}
\usepackage[english]{babel}
\usepackage{amsfonts}
\usepackage{mathrsfs}
\usepackage{amsthm}
\usepackage{amssymb}
\usepackage{enumerate}
\usepackage{amsmath}
\usepackage{subfigure}
\usepackage{algorithm}
\usepackage{algorithmicx}
\usepackage{algpseudocode}
\usepackage{hyperref}

\usepackage{multirow}

\jyear{2021}%

\theoremstyle{thmstyleone}%
\theoremstyle{thmstyletwo}%

\theoremstyle{thmstylethree}%

\begin{document}

\title[Article Title]{The Dynamic Resilience of Urban Labour Networks}


\author[1]{\fnm{Xiangnan} \sur{Feng}}\email{fengxiangnan@gmail.com}
\equalcont{These authors contributed equally to this work.}

\author*[1]{\fnm{Alex} \sur{Rutherford}}\email{alexisadams@gmail.com}
\equalcont{These authors contributed equally to this work.}

\affil[1]{\orgdiv{Centre for Humans and Machines}, \orgname{Max Planck Institute for Human Development}, \orgaddress{\street{Lentzeallee 94}, \city{Berlin}, \postcode{14195}, \country{Germany}}}


\abstract{Understanding and potentially predicting or even controlling urban labour markets represents a great challenge for workers and policy makers alike. Cities are effective engines of economic growth and prosperity and incubate complex dynamics within their labour market, and the labour markets they support demonstrate considerable diversity. This presents a challenge to policy makers who would like to optimise labour markets to benefit workers, promote economic growth and manage the impact of technological change. While much previous work has studied the economic characteristics of cities as a function of size and examined the exposure of urban economies to automation, this has often been from a static perspective. In this work we examine the structure of city job networks to uncover the diffusive properties. More specifically, we identify the occupations which are most important in promoting the diffusion of beneficial or deleterious properties. We find that these properties vary considerably with city size.}




\maketitle

\section{Introduction}\label{sec1}

Cities and work are closely intertwined, with many cities historically converging around locations advantageous for commerce, leading to economies of scale and increasing returns~\cite{Bettencourt7301, pinheiro2021time}, and more recently encouraging the urban migration of workers to convene in centralised workplaces~\cite{Glaeser_2012, toth2021inequality}. However, setting labour policy on an urban level (or indeed nationally or regionally) in an optimal way is both critical for prosperity yet challenging on a technical level~\cite{ferragina2022labour}. Labour markets, the allocation of workers with skills to tasks required to be performed by organisations, are extremely complex. In order to better understand this complexity, much work has begun to represent labour markets, skills and jobs as complex networks~\cite{alabdulkareem2018unpacking,r_maria_del_rio_chanona_2020_4453162,dworkin,guerrero2013employment,schweitzer2009economic,park2019global,meindl2021four}.

Networks are a natural representation for, not only how workers move between jobs that are mutually accessible based on skills, but also for diffusive processes by which technology spreads~\cite{UBHD2028615}. Much, although not all, previous work has focused on static properties of the labour networks. However labour markets are dynamic with flow of workers between jobs, of working practices between organisations and automation within firms and industrial sectors. Therefore, in this work we focus on the degree of resilience in an urban job network based on its network structure amenable to diffusion. This view is consistent with a large body of work examining the role of relatedness in the study of economic complexity~\cite{hidalgo2021economic}

From the perspective of a policy maker, there are a number of attributes that might be expected to diffuse on a network of occupations. Some of these might be optimal e.g. a gender balanced work force or safer and healthier working conditions, while some might be unfavourable e.g. skill based technological change leading to the displacement of human labour. However, with knowledge of the degree to which a local job network will be able to promote or constrain diffusion and the nodes which are most influential in this network, policy makers can more effectively drive the labour market to a desired state. For example, the adoption of company wide policies protective of children and young people has been found to preferentially occur along network connections defined by supply chains~\cite{labour_policies}.

\begin{figure}[htbp]
	\centering
	\includegraphics[width=0.9\textwidth]{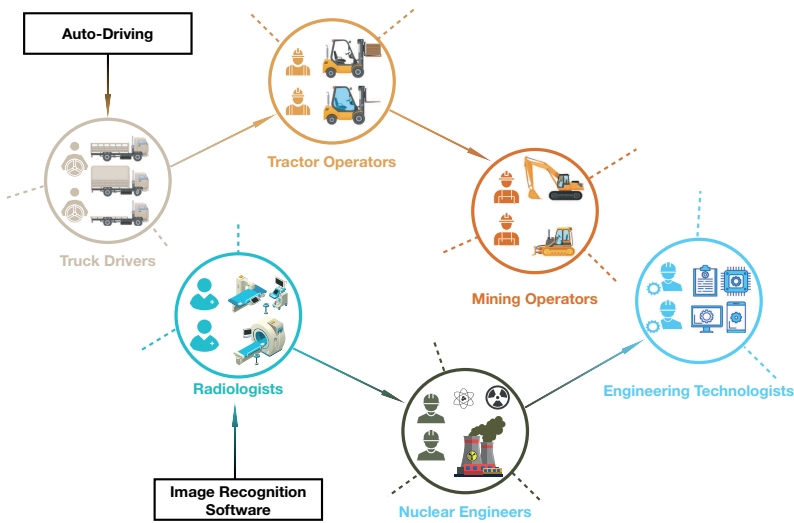}	
	\caption{An illustration of workplace technologies spreading on an occupational network. Autonomous Driving capability is first introduced for Truck Drivers, which subsequently spreads to impact Tractor Operators and then Mining Operators due to the similarities between these occupations. Likewise Image Recognition Software introduced for Radiologists spreads to Nuclear Engineers and Engineering Technologists. Not all occupations are drawn in this diagram; only those lying on one illustrative diffusive pathway.}\label{spreadingexample}
\end{figure}

In this manuscript, we focus on the mechanisms by which new technology diffuses (although this is amenable to the diffusion of other occupation based norms). As shown in Fig \ref{spreadingexample}, when the workplace tasks of truck drivers are exposed to technologies such as autonomous-driving systems, this automation could spread preferentially to a related job e.g. tractor operators. Subsequently other occupations similar to tractor operators with respect to skills could also be effected. However, this phenomenon is not limited to physical jobs, image recognition software can assist radiologists diagnosing diseases in medical imagery. This same technology could be learned and adopted by nuclear engineers to detect the operation of nuclear power plants; then engineering technologists could apply this software onto more instruments to gain more powerful tools. 

We wish to emphasise that the spread of technology might be beneficial or deleterious depending on the technology itself, as well as the stakeholder. A piece of technology might displace human labour completely leading to unemployment, but benefit a firm which is able to lower its wage bill. Likewise, a new piece of technology might well enable increased productivity in occupations depending on the efficiency of the technology in question~\cite{goldfarb2019economics}.

Our contributions in this work are focused on providing a novel understanding of the structure and resilience of urban labour markets. More specifically we (i) propose the employment weighted spreading influence, a new measure of a node's influence in a job network (ii) we quantify the efficiency of diffusion based on a seeding strategy on this basis (iii) we investigate how the cognitive and automatable nature of influential jobs changes with the size of the urban economy and (iv) we uncover the most influential occupations; those that would optimise the efficacy of targeted policy interventions, across national urban centres.

\section{Data}
In this research, we consider data from O*NET~\cite{onet} and U.S. Bureau of Labor Statistics~\cite{bls}. Although similar structured data is increasingly available e.g. in the EU~\cite{esco}, the use of O*NET is common in network studies of labour, and given the high quality of this data broken down by urban area, we proceed with these data. From this structured data, we build networks for each urban area as well as nationally. All the analyses are based on the networks.

The O*NET data contains a set of standardised occupations and for each occupation, a weighted importance against a set of standard skills. This data is compiled manually from expert interviews based on the skills that workers report. We consider the data from 2011 and 2020 as the Standard Occupational Classification (SOC) system is consistent in this time. This includes 774 occupations globally and 120 items referred as ``skills'' including ability, knowledge and skill sections in the dataset are used to build networks. We use 2-digit and 6-digit SOC classifications for analysis.

The U.S. Bureau of Labor Statistics provides employment data including occupations, employment numbers and salaries in each Metropolitan Statistical Area, commonly referred to as a ``city''. Around 380 cities are selected in the analysis. Not all occupations appear in every city, ranging from 93 to 733 occupations in different areas. We build networks for each city based on the occupations found in each city (as described in more detail below).

Other data includes population data from U.S. census~\cite{census} and occupation automation probability data from Frey and Osborne's and Michael Webb's research~\cite{frey2017future, webb2019impact}. The occupations in different data sources are indexed by SOC codes.

\section{Methods}
\subsection{Network Construction}
In the created occupation networks, each node stands for one occupation and edge stands for similarity between the two occupations whose weight is given by skill similarity. We use the Jaccard similarity~\cite{jaccard1912distribution} to quantify the similarity between occupation $i$ and $i'$:
\begin{eqnarray}
SkillSimilarity(i,i') = \frac{\sum_{s}\min(onet_{i,s}, onet_{i',s})}{\sum_{s}\max(onet_{i,s}, onet_{i',s})},
\end{eqnarray}
where $onet_{i,s}$ is the importance of skill $s$ to occupation $i$. High skill similarity between jobs suggests workers may more easily transition between them. 

Since this similarity value between any occupation pair is not zero, a complete network will be built containing 774 nodes and $299,151$ edges, which is the global network composed of all occupations in U.S. In each city, its occupation network is built in the same way with occupations whose employment is non-zero locally, which means each local occupation network is a complete sub-network of the global one. 

\subsection{Employment Spreading Influence}
With a view of understanding the dynamic behaviours among various individuals on networks, a number of theories studying the interactions between nodes have been considered. These operate globally and locally to model the spreading processes, which traditionally usually focus on information~\cite{pei2014searching} and epidemic spread~\cite{kang2020spatial}. A key issue is to find the ``super-spreader'', namely the nodes with highest spreading abilities. Controlling these super-spreaders, that is purposefully seeding with them or attempting to isolate them, can lead to optimal spreading or immunization respectively~\cite{kitsak2010identification}.

The spreading also happens on the labour markets and occupation networks. New technologies and skills from one occupation can potentially diffuse into other occupations, as since new uses cases beyond the original promotes the introduction and adaptation of existing inventions. For example, the recording of speech was firstly invented by Edison to record the last words before people dying and books for blind persons, then it was adapted for office work. But even Edison himself did argue against to use this invention to record music until around 20 years after, which totally revolutionized the whole music industry. If we view the skill similarity based network from the perspective of spreading, the super-spreaders, namely the most influential occupations, would present the greatest ability to spread technologies or norms to the whole labour market. These occupations play crucial roles in leading and shaping the landscape of labour market. On one hand, if starting with them, new technologies improving efficiency could be adopted faster, which increases the  productivity and economic growth. On the other hand, automation of skills and jobs could spread, leading to potentially negative effects such as short-term displacement of workers~\cite{aiwork}.

From this perspective, if we could obtain the most influential occupations, it could help regulate the whole labour market. For policy makers, understanding which occupations are influential and making policies targeting at these occupations and skills could lead the labour market to be robust leading to favourable economic outcomes.

Finding the optimal spreaders belongs to the NP-hard~\cite{kempe2003maximizing, karp1972reducibility} problems generally. Interactions at different topological levels need to be studied~\cite{rosvall2014memory, holland1977method} and multi-scale features tangling together increases the difficulty. It was studied that high influence nodes are not always those whose degrees are highest~\cite{morone2015influence}. By applying the cavity method~\cite{mezard2003cavity} on this weighted network case~\cite{zhang2018dynamic}, we define the spreading influence of occupation $i$ in city $c$:
\begin{eqnarray}
SI_i^c = \sum_{j\in N(i)}a_{ij}\sum_{k\in N(i),k\ne j}a_{ik}(d_j-1),
\end{eqnarray}
where $a_{ij} = SkillSimilarity(i,j)$ is the edge weight between node $i$ and $j$, $d_j$ is the node degree. The derivation could be seen in Supplementary Information Section~\ref{derivation}.

Besides the network structure and skill similarity, the number of workers to be found at each occupational node of the network is highly inhomogeneous which has a strong influence on the labour market as a whole. Occupations with more employees should always be preferentially targeted by policies, since when new technologies applied, they will have a large impact on the local or global labor market, either in good way or in a bad way to to instability. Thus, when measuring the influence of occupations, we take into account the employment of occupation $i$ in city $c$, denoted as $emp_i^c$, to get the employment spreading influence:
\begin{eqnarray}
empSI_i^c &&=\log(emp_i^c)SI_i^c\\
&&= \log(emp_i^c)\sum_{j\in N(i)}a_{ij}\sum_{k\in N(i),k\ne j}a_{ik}(d_j-1).\nonumber
\end{eqnarray}

For each city, the labour market is described by its occupation network and the $empSI$ of each occupation in that city is calculated. Occupations with high $empSI$ suggest high influence on the local market in the perspective of spreading, which has a contribution due to both their structural spreading abilities or large employment numbers. For policy makers at each level, these occupations deserve particular emphasis.

\section{Results}
In this manuscript, we investigate the structure of the labour market in each city. More specifically, we wish to identify the most influential occupations in each city network and to determine how this relates to city size. For each city $c$, the $empSI_i^c$ values of all occupations that are represented in the city workforce are calculated and those occupations with highest values are considered the most influential in a given city. 

\begin{figure}[htbp]
	\centering
	\includegraphics[scale=0.23]{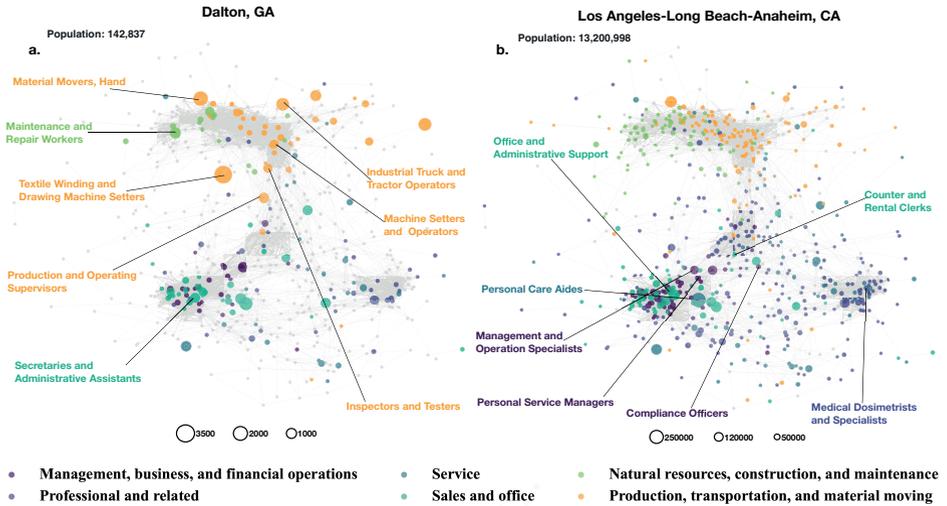}	
	\caption{Two occupation networks of \textbf{(a.)} Dalton, GA and \textbf{(b.)} Los Angeles-Long Beach-Anaheim, CA. Nodes stand for occupations and edges stand for skill similarity between occupations. Node colors stand for sectors and sizes indicate the employment numbers. The gray node indicates that this occupation does not appear in this city. Edges with weights higher than 0.7 are drawn, while the complete graph is applied in calculating. The labelled occupations are the different occupations in the two cities' top-20 highest $empSI$ occupations. }\label{networks}
\end{figure}

To begin, we compare the job networks of two cities of very different sizes; Dalton, GA (low population) and Los Angeles, CA (high population) in Figure \ref{networks}. The networks link nodes (occupations) by skill similarity (edges). We consider only the jobs that are found in each city, therefore the set of occupations (number of nodes) is typically different between cities. Each occupation has a variable number of workers performing that occupation and is represented by the size of the corresponding node. A selection of the most influential occupations are marked in each case. A distinction can be made between the two networks: in Dalton the most influential occupations concentrate on the production and construction sectors, while in LA, the most influential ones belong to services, sales and clerical sectors. 

\begin{figure}[htbp]
	\centering
	\includegraphics[scale=0.28]{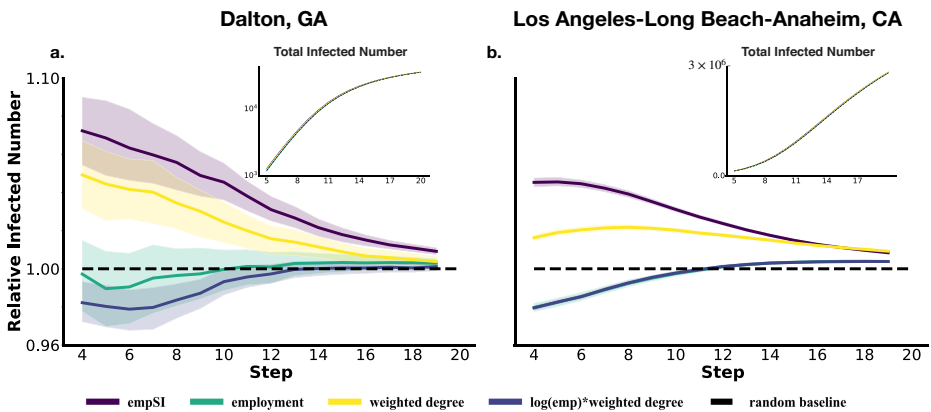}	
	\caption{Simulation of a spreading process on the occupation networks of \textbf{(a.)} Dalton, GA and \textbf{(b.)} Los Angeles-Long Beach-Anaheim, CA. All results are the averaged values of 20 trials. In each simulation, $0.2 \%$ random agents of employment are selected by different strategies including highest $empSI$, highest employment numbers, highest weighted degrees and highest $\log{(emp)}\times$ weighted degree as the starting seeds for spreading. Each node in the networks contains the same agent number as employment. At each step, each agent infects one of its neighbours with the probability given by their skill similarity. A strategy of random seed selection is applied as a baseline. The curves are the relative infected number and inside is the total infected number. }\label{spreading}
\end{figure}

To verify whether $empSI$ meaningfully represents influential occupations, a simple spreading simulation within the Susceptible-Infected framework is conducted on one local network, for the two areas above (See model details in Supplementary Information Section~\ref{simulation}). Figure \ref{spreading} presents the spreading processes starting with seed agents selected by different strategies. The random selection of the seeds is applied as a null model. As shown in the Figure, seeding based on the $empSI$ value, leads to a faster spread than seeding at random or based only on employment or degree.

\subsection{Diversity and Automation}

Next we investigate the relationship between the characteristics of an occupation; the cognitive nature and exposure to automation, and the occupations importance within a city job network.

\begin{figure}[htbp]
	\centering
	\includegraphics[scale=0.3]{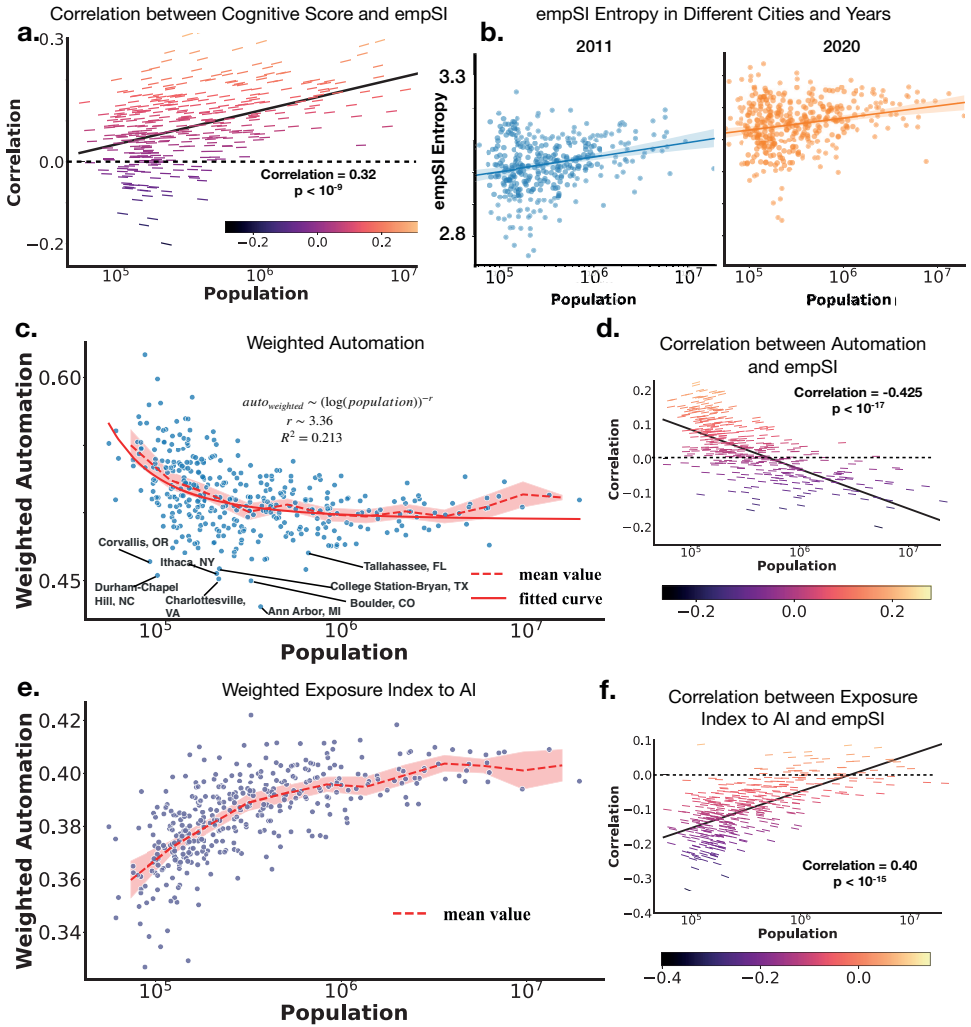}	
	\caption{\textbf{(a.)} Correlations between cognitive score values and $empSI$ against populations in each city. The Pearson correlation coefficient is marked. Different colours and slopes stand for different correlation values.\textbf{(b.)} The $empSI$ entropy of year 2011 and 2020 against populations. For each city, occupation $empSI$ values are divided into same number of bins and its entropy is calculated. \textbf{(c.)} Weighted automation probability by $empSI$ against populations. The mean value is the average values in each bin which is log-scale by populations. A log-power fitting is implemented on the data. On the left bottom, a group of college towns are labelled, featured by small-scale population and low weighted automation probability. \textbf{(d.)} Correlations between automation and $empSI$ against population. \textbf{(e.)} Weighted exposure index to AI from Webb's research by $empSI$ against populations. \textbf{(f.)} Correlations between exposure to AI and $empSI$ against population.}\label{empsicor}
\end{figure} 

For each city, we compute the correlation between all occupations' $empSI$ values and cognitive scores, which is defined as the fraction of cognitive skills for each occupation~\cite{alabdulkareem2018unpacking}, namely $\sum_{s\in cognitive}onet_{i,s}/\sum_{s}onet_{i,s}$. In smaller cities we find little correspondence between the two measures; in this case, the more influential occupations in small cities are just as likely to be cognitive or non-cognitive. Whereas in larger cities we find a reasonable correlation ($\rho \approx 0.3$) between these measures. This is demonstrated by a clear positive trend between city size and the job-wise correlation in cognitive score and $empSI$ (r,p) = (0.32, $<10^{-9}$). Thus we expect that any occupational characteristic present in more cognitive jobs, whether positive or negative, will spread more efficiently to other occupations in larger cities. Interestingly, in contrast to many economic indicators~\cite{Bettencourt7301}, we found no significant correspondence to a super- or sub-linear scaling with population.

We note that this trend of increasing influence of cognitive jobs with city size also coincides with an increasing diversity of job influence, both with respect to city size and over time. We calculate the Shannon entropy~\cite{shannon1948mathematical} of $empSI$ in each city. Figure \ref{empsicor}b demonstrates that the entropy of node influence across all occupations found in a city increases with population. Further, the level of diversity appears to be generally increasing between 2011 and 2020. Taken together, these results suggest that in larger cities, the correlation between the cognitive nature of jobs and their influence is not driven by a small number of highly influential cognitive jobs. Rather, the increasing uniformity in $empSI$ both over time and with city size, suggests that the choice of occupations for effective targeted policy interventions is not trivial (see a similar result related to $SI$ and cognitive score in Supplymentary Information Section~\ref{cor_si_cs}).

Next we consider whether these trends in increasing diversity and more influential cognitive jobs in larger cities lead to increased \textit{resillience} to automation specifically. In Figure \ref{empsicor}c we evaluate the city-wise exposure to automation as measured by Frey and Osborne~\cite{frey2017future}, weighted by $empSI$ as below
\begin{eqnarray}
auto_{weighted}^c = \sum_i \frac{empSI_i^c}{\sum empSI}\times P_{auto}(i),
\end{eqnarray}
where $P_{auto}(i)$ is the automation probability of occupation $i$ from Frey and Osborne~\cite{frey2017future}.

In common with previous findings~\cite{frank2018small}, we see a decreasing exposure to automation with city size when weighted by $empSI$: smaller cities face a greater risk of automation. At the same time, at the left-bottom part of the figure, we observe a group of outliers, mostly corresponding to college towns including Ithaca, NY (Cornell University), Ann Arbor, MI (University of Michigan), Charlottesville, VA (University of Virginia) and Durham-Chapel Hill, NC (Duke University and North Carolina Central University). This phenomenon is consistent with the intuition that these college towns should be under low automation risk.

Further, the occupations that are most exposed to automation are less influential in larger cities. This result goes beyond a static picture of a city's overall automation risk as measured by its present workforce. Rather this index measures the degree to which an occupation is able to spread its own exposure to automation to other related occupations. Although this trend is relatively weak (from $\rho \approx 0.2$ in the smallest cities to $\rho \approx -0.2$ in the largest cities) it is consistent across all cities ($(\rho,p) = (-0.425, < 10^{-17})$).

However we find that this result is extremely sensitive to the exact measure of automation exposure. In Figure \ref{empsicor}d, we present the same weighted automation measure using data from~\cite{webb2019impact}, an index of exposure to AI specifically.  In contrast, we find the opposite trend in the relationship between the urban exposure to automation and population. Specifically, the weighted automation exposure has a positive correlation with city size ($(\rho,p) = (0.40, < 10^{-17})$). Similar calculations are repeated based on other automation data~\cite{webb2019impact, brynjolfsson2018can}, see in Supplementary Information Section~\ref{wauto}.

The exposure of an occupation to automation is notoriously difficult to quantify, and the measures presented here all differ in precisely what is measured. Here we principally compare Frey and Osbourne's results as the most established measure on the one hand, and Webb's data as a more recent measure based on the similarity between occupation task descriptions and patent documents. Webb's measure offers the key advantage of being validated on historical changes in workforce numbers, and is also broken down into the categories of AI, software and robotics. 

Our results suggest that the cognitive jobs found preferentially in larger cities are more amenable to automation through the deployment of AI technology (as opposed to robots or software). AI technology can be more easily deployed in work environments where computers and data infrastructures are already common place as well as through flexible cloud computing resources. The fact that no correspondence is found between city size and equivalent measures of automation exposure through robots and software support this. 

These results demonstrate that urban labour market resilience has a nuanced relationship with city size and depends sensitively on the nature of the occupational automation risk. We can conclude that larger cities are more exposed to AI based automation technologies and that cognitive jobs will be able to diffuse occupational characteristics, possibly through targeted policy interventions, more efficiently.

\subsection{Landscape of USA Labour Markets}
Previous works has shown that the industry structure varies geographically~\cite{park2019global, molinero2021geometry, chen2021automation}. For different cities, is there any pattern or particular combination of the influential occupations? What do the most common influential occupations and uncommon ones look like across cities? 

\begin{figure}[htbp]
	\centering
	\includegraphics[width=0.9\textwidth]{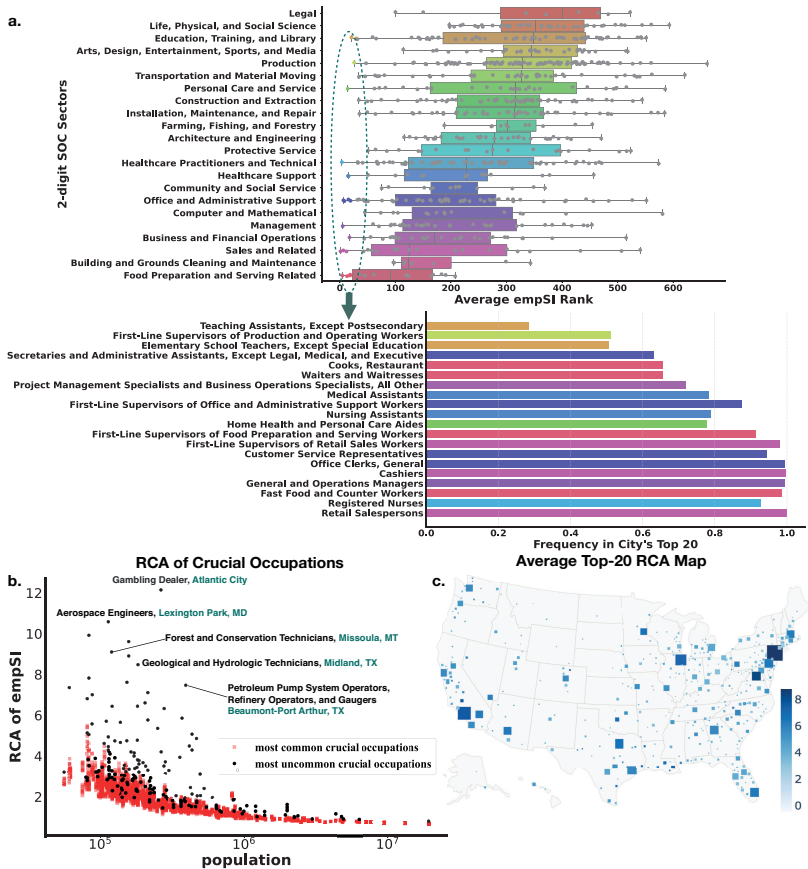}	
	\caption{\textbf{(a.)} Box plot of average $empSI$ ranks of occupations in each sector. The two-digit SOC code is applied to divide occupations into different sectors and different colors stand for different sectors. For each occupation, its $empSI$ in every city is calculated and the average $empSI$ rank is calculated. We select the 20 highest average rank occupations (marked by colourful nodes in the box plot) to count their frequencies of appearing in each city's top-20 $empSI$. \textbf{(b.)} RCA values of most common crucial occupations and uncommon crucial occupations against populations. The crucial occupations are defined as the top-20 $empSI$ ones in each city. Several occupations are labelled.\textbf{(c.)} The map of top-20 $empSI$ RCA occupations in all cities. Square size stands for populations and colour stands for the value.}\label{rcarank}
\end{figure} 

To find the common pattern of influential occupations, in each city we rank all the active occupations by their local $empSI$ and for each occupation the average rank is calculated. We investigate the occupations with highest average $empSI$ ranks by counting their frequency of ranking Top 20 in local labour markets. As shown in Figure \ref{rcarank}\textbf{a}, we find that different sectors demonstrate different patterns of which occupations are found in higher ranks; there are some occupations that are crucial and influential in almost all the cities and most of them concentrate in sales, services, health and office sectors, like Retail Salespersons, Registered Nurses, Cashiers and Office Clerks. These occupations, due to both large employment numbers or high spreading abilities based on skill similarities, play influential roles across all cities. It could be observed that most of these occupations are regarded as basic life-supporting and economic-activity ones. For policy makers, these occupations are crucial since they are fundamental and prevailing for local economic, and global policies targeting these occupations may cause similar effects around the whole country.

While there are a group of common influential occupations across the whole nation, one might wonder what makes the labour market particular to the cities? Previous work has shown that features vary with the scale of cities, including wages, patent activity and characteristic industries~\cite{david2015there, acemoglu2018race, hong2020universal}. Are the existing influential occupations specific to particular cities? To verify this, we calculate the revealed comparative advantage (RCA)~\cite{hidalgo2021economic} of occupation $empSI$ in each city:
\begin{eqnarray}
{RCA}_i^c = \frac{empSI_i^c / \sum_i empSI_i^c}{\sum_c empSI_i^c/\sum_i\sum_cempSI_i^c}.
\end{eqnarray}
This index could help find the relative advantage of an occupation, namely occupations with high $RCA_i^c$ will present high influence on local labour market and this influence is identical to this city -- in cities other than $c$, these occupations are either playing negligible roles or not active. We present $RCA_i^c$ values of some most common and uncommon crucial occupations in Figure \ref{rcarank}\textbf{b}. It could be observed that there are a number of occupations featured by high $RCA_i^c$ being influential in only one or several places, like Gambling Dealer in Atlantic City which is famous for gambling industry, Aerospace Engineers in Lexington Park in Maryland where the NASA Goddard Space Flight Center is close, Geological and Petroleum related work in Midland and Port Arthur in Texas where the oil sources were discovered. These high $RCA_i^c$ occupations present strong geographical identities (see in Figure \ref{rcarank}\textbf{c}) and it could be observed in large cities like New York and Los Angeles, their average $RCA_i^c$ values of top influential occupations are high, suggesting there are more occupations being crucial in these areas. These geographically identical occupations deserve more attention from policy makers, since they are rare and could outcome special productions across the whole nation.

In summary, it has been shown from this $empSI$ perspective that there are some occupations that appear and play crucial roles in every city, which have high average $empSI$ ranks and usually focus on life and health sectors, like sales, services, health and office. They are supporting the everyday routine life in the global nation and prevailing in many professions. At the same time, there are some unique occupations in cities, which are deeply related to the local economy structures and city scales like college towns and nature resource centres, being crucial and influential locally. They tend to have high $empSI$ RCA values and deserve specialised attention and policies. These two aspects of crucial occupations, the globally crucial ones and locally crucial ones, make up the backbones of U.S. labour market landscape.

\section{Conclusion and Discussion}

The adoption of workplace behaviours, like many other social norms, can have positive and negative effects. In the context of workplace norms, the adoption of safer working conditions would be beneficial for workers whereas the adoption of new technology could potentially be a negative development for both workers and policy makers. Given the many successes of describing complex social systems, such as labour markets, as complex networks it is natural to consider the diffusive properties of labour markets. This provides an attractive opportunity for policy makers to take advantage of this diffusive property to encourage (inhibit) the faster (slower) adoption of positive (negative) attributes on the network of occupations.

In the perspective of spreading, we have proposed an employment spreading index to measure the occupation influence on occupation networks. For each city, an occupation network is built based on skill similarities and the most influential occupations are obtained by the index. We investigate the systematic effect of city size on the susceptibility of jobs. We find that cognitive jobs are consistently more influential as city population increases. Regarding the effect of automation specifically, we find that the relationship with occupational influence is more nuanced. The Frey and Osbourne measure of exposure to automation suggests that larger cities will be more resillient to automation as the influential occupations are those with a lower exposure. Converseley, Webb's measure of exposure to AI specifically suggests the opposite trend; larger cities have their more influential occupations more exposed.

Considering influential occupations on a global level, by aggregating the most influential occupations across all cities, we are able to rank occupations. We find that the top jobs tend to include an element of physical and socio-cognitive work e.g. Nurses, Cashiers and Customer Service Representatives. These results suggest that these mixed-nature jobs should be given special attention by policy makers when considering how to manage labour market change whether technological in nature or not.

In this work we have considered the network of occupations linked by the similarity of the tasks required. However behaviours such as the adoption of automation technologies are able to diffuse on many network substrates; between the successive jobs of individual workers, between firms in common sectors in physical space or along supply chains. Considerable explanatory power has been found in novel combinations of these e.g. geo-industrial networks~\cite{park2019global} and supply chain networks~\cite{lafond2022reconstructing}. It is likely that the true dynamics involve some combination of these into a multiplex network\cite{tuninetti2021prediction}, on which the cooperative spreading of more than one social norms would diffuse~\cite{mivsic2015cooperative, hyland2021multilayer}. More related studies of the diffusion mechanism on labour market networks could bring deeper understanding on them. A key challenge will be how to apply new strategies to capture these patterns, like research on high-order interactions~\cite{battiston2021physics, millan2021local} and structures including motif~\cite{benson2016higher, milo2002network} and graphlet~\cite{sarajlic2016graphlet}. The occupation network structure deserves more explorations in the future.

Our study is constrained by the availability of public data. Higher fidelity data would allow for more careful validation of these findings. We also make conceptual simplifications that the occupational network is static of the timescale for diffusion to take place. Implicitly this means that we ignore the fact that the labour market is dynamic: new occupations are generated every year and for many occupations new skills and working contents are introduced~\cite{acemoglu2019automation, borner2018skill}. 

\section{Acknowledgements}
We thank Dr Manuel Cebrian for numerous constructive conversations while developing this project.

\begin{appendices}

\section{Derivation of Spreading Influence}\label{derivation}
This result by 1RSB is assumed on sparse network~\cite{mezard2003cavity, zhang2018dynamic}, especially on locally tree-like networks where there are not many loops, yet it still could works on more dense networks. In the research we applied the weighted complete networks to model the occupation networks and the conclusion still works. In real world case, the labour market could be less connected and it is possible the network will not be complete, which suggests that this conclusion may work even better in real world cases.

Under the sparse network assumption, for a network $G$ with node set $V$ and edge set $E$, its adjacency matrix $A$ is defined as
\begin{eqnarray}
A_{ij}=\left\{
\begin{array}{ll}
1 & \textrm{if $v_i$ and $v_j$ are connected}\\
0 & \textrm{otherwise.}
\end{array}\right.
\end{eqnarray}
In a weighted network case, we use $a_{ij}$ to stand for the weight of edge between node $v_i$ and $v_j$.

The maximum spreading problem equals to the optimal percolation problem~\cite{morone2015influence}, which is how to break down the major component in networks with minimal set of nodes removed. Let $c_{i\to j}$ denote the probability of node $v_i$ belonging to the major component with node $v_j$ removed. For a tree-like weighted network, this relation could be formulated as:
\begin{eqnarray}
c_{i\to j} = n_i[1-\prod_{h\in \partial i, h\neq j}(1-c_{h\to i}a_{ih}A_{ih})],
\end{eqnarray}
where $n_i = 0$ stands for node $v_i$ being an optimal spreader and not if $n_i = 1$and $h\in \partial i$ means the neighbours of node $v_i$.

Based on knowledge from dynamic system, the above system will have a stable solution if the largest eigenvalue of linear operation $\mathop{R}$ is smaller than 1~\cite{bhatia2002stability}. Here, the $\mathop{R}$ is a $2\vert E\vert \times2\vert E\vert$ matrix and each row and column corresponds to a one-direction edge. The matrix $\mathop{R}$ is defined as:
\begin{eqnarray}
\mathop{R_{i\to j, k\to l}} = \frac{\partial c_{i\to j}}{\partial c_{k\to l}}\vert_{c_{i\to j=0}},
\end{eqnarray}
where $i\to j$ and $k\to l$ stand for the edges between $v_i$ and $v_j$ and $v_k$ and $v_l$ with directions. $\mathop{R}$ could be calculated by non-backtracking matrix $\mathop{M}$~\cite{hashimoto1989zeta}:
\begin{eqnarray}
\mathop{R_{i\to j, k\to l}} = n_k a_{ij} \mathop{M_{i\to j, k\to l}},
\end{eqnarray}
where
\begin{eqnarray}
\mathop{M_{i\to j, k\to l}} = \left\{
\begin{array}{ll}
1 & \textrm{if $i=j$, $k\neq l$}\\
0 & \textrm{otherwise}.
\end{array} \right.
\end{eqnarray}
Thus, we have:
\begin{eqnarray}
\mathop{R_{i\to j, k\to l}} = n_k a_{ij} A_{ij} A_{kl}\delta_{jk}(1 - \delta_{il}),
\end{eqnarray}
where:
\begin{eqnarray}
\delta_{ij}=\left\{
\begin{array}{ll}
1 & \textrm{if $i=j$}\\
0 & \textrm{otherwise}.
\end{array} \right.
\end{eqnarray}
In the equation above, $A_{ij}$, $A_{kl}$ and $\delta_{jk}$ suggest that there is a path $v_i\to v_j \to v_k \to v_l$ and $\delta_{il}$ make sure that it is a non-backtracking path. 

Let $\lambda({\mathbf{n}})$ be the largest eigenvalue of $\mathop{R}$ where $\mathbf{n} = (n_1,n_2,\dots,n_{\vert V\vert})$ indicates which nodes are selected as optimal spreader. We apply the power method~\cite{atkinson2008introduction} from numerical analysis to approximate $\lambda({\mathbf{n}})$. Let $\mathbf{w}_0$ be a vector satisfying that it has nonzero projection on the direction of $\lambda({\mathbf{n}})$'s eigenvector and let $\textbf{w}_r(\textbf{n})$ be:
\begin{eqnarray}
\textbf{w}_r(\textbf{n})=\mathop{R^r}\textbf{w}_0.
\end{eqnarray}
Then, according to power method, we could get:
\begin{eqnarray}
\lambda(\textbf{n})=\lim_{r\to \infty}\lambda_r(\textbf{n})=\lim_{r\to \infty}\bigl(\frac{\vert \textbf{w}_r(\textbf{n})\vert }{\vert \textbf{w}_0\vert }\bigr)^{\frac{1}{r}}.
\end{eqnarray}
where 
\begin{eqnarray}
\vert \textbf{w}_r(\textbf{n})\vert ^2 = \langle \textbf{w}_r(\textbf{n})\vert \textbf{w}_r(\textbf{n})\rangle=\langle \textbf{w}_0\vert (\mathop{R^r})^T\mathop{R^r}\vert \textbf{w}_0\rangle.
\end{eqnarray}

When $r = 1$, the approximated eigenvector $\mathbf{w}_1$ is:
\begin{eqnarray}
\vert \textbf{w}_1(\textbf{n})\rangle = \mathop{R}\vert \textbf{w}_0\rangle.
\end{eqnarray}

Let $\vert \textbf{w}_0\rangle = \vert \textbf{1}\rangle$, the left vector could be calculated as:
\begin{eqnarray}
_{i\to j}\langle\textbf{w}_1(\textbf{n})\vert  &&= \sum_{k\to l}{_{k\to l}}\langle\textbf{w}_0\vert \mathop{R_{k\to l,i\to j}}\nonumber\\
&&= \sum_{k\to l}n_i a_{kl}A_{kl}A_{ij}\delta_{li}(1-\delta_{kj})\nonumber\\
&&=n_i A_{ij}\sum_{k\to l}a_{kl}A_{kl}\delta_{li}(1 - \delta_{kj}),
\end{eqnarray}
and the right vector:
\begin{eqnarray}
\vert \textbf{w}_1(\textbf{n})\rangle_{i\to j} && = \sum_{k\to l}\mathop{R_{i\to j,k\to l}}\vert \textbf{w}_0\rangle_{k\to l}\nonumber\\ 
&& = \sum_{k\to l} n_j a_{ij} A_{ij} A_{kl} \delta_{jk} (1 - \delta_{il})\nonumber\\
&& = n_j a_{ij} A_{ij} \sum_{k\to l} A_{kl} \delta_{jk} (1 - \delta_{il})
\end{eqnarray}
So the norm of $\textbf{w}_1(\textbf{n})$ is :
\begin{eqnarray}
\vert \textbf{w}_1(\textbf{n})\vert ^2 && = \sum_{i\to j}{_{i\to j}}\langle\textbf{w}_1(\textbf{n}) \vert  \textbf{w}_1(\textbf{n})\rangle_{i\to j}\nonumber\\
&&= \sum_{i\to j}n_i n_j a_{ij} A_{ij} \times \nonumber\\
&& \sum_{k\to l}a_{kl}A_{kl}\delta_{li}(1 - \delta_{kj}) \sum_{k\to l}A_{kl}\delta_{li}(1 - \delta_{kj})
\end{eqnarray}
Let $M_{i\to j} = \sum_{k\to l} a_{kl}A_{kl}\delta_{li}(1 - \delta_{kj})$ and $N_{i\to j} = \sum_{k\to l}A_{kl}\delta_{jk} (1 - \delta_{il})$, then 
\begin{eqnarray}
\vert \textbf{w}_1(\textbf{n})\vert ^2 = \sum_{i\to j} n_i n_j a_{ij} A_{ij} M_{i\to j} N_{i\to j}
\end{eqnarray}
and the approximation to largest eigenvalue is:
\begin{eqnarray}
\lambda_1(\textbf{n}) = \bigl(\frac{1}{2\vert E\vert } \sum_{i\to j} n_i n_j a_{ij} A_{ij} M_{i\to j} N_{i\to j}\bigr)^{\frac{1}{2}}.
\end{eqnarray}

For $r=2$, the left vector $_{i\to j}\langle\textbf{w}_2(\textbf{n})\vert $ and right vector $\vert \textbf{w}_2(\textbf{n})\rangle_{i\to x=j}$ could also be calculated:
\begin{eqnarray}
_{i\to j}\langle\textbf{w}_2(\textbf{n})\vert  && = \sum_{k\to l}{_{k\to l}}\langle\textbf{w}_1\vert \mathop{R_{k\to l, i\to j}}\nonumber\\
&& = \sum_{k\to l}n_i n_k a_{kl} A_{kl} A_{ij} \delta_{li} (1 - \delta_{kj}) M_{k\to l}\nonumber\\
&& = n_i A_{ij} \sum_{k\to l} n_k a_{kl} A_{kl} \delta_{li} (1 - \delta_{kj}) M_{k\to l};
\end{eqnarray}
\begin{eqnarray}
\vert \textbf{w}_2(\textbf{n})\rangle_{i\to j} && = \sum_{k\to l}\mathop{R_{i\to j,k\to l}}\vert \textbf{w}_1\rangle_{k\to l}\nonumber\\
&& = \sum_{k\to l} n_l a_{kl} A_{kl} N_{k\to l} n_k a_{ij} A_{ij} \delta_{jk} (1 - \delta_{li})\nonumber\\
&& = a_{ij} A_{ij} \sum_{k\to l} n_k n_l a_{kl} A_{kl} \delta_{jk} (1 - \delta_{li}) N_{k\to l}.
\end{eqnarray}
Thus, the second order eigenvector is
\begin{eqnarray}
\vert \textbf{w}_2(\textbf{n})\vert ^2 && = \sum_{i\to j}{_{i\to j}}\langle\textbf{w}_1(\textbf{n}) \vert  \textbf{w}_1(\textbf{n})\rangle_{i\to j}\nonumber\\
&& = \sum_{i\to j, k\to l} n_i n_j n_k n_l a_{ij} a_{ki} a_{lj} A_{ij} A_{ki} A_{lj} M_{k\to i} N_{j\to l}
\end{eqnarray}
and the approximated eigenvalue is
\begin{eqnarray}
\lambda_2(\textbf{n}) = \bigl(\frac{1}{2\vert E\vert } \sum_{i\to j, k\to l} n_i n_j n_k n_l a_{ij} a_{ki} a_{lj} A_{ij} A_{ki} A_{lj} M_{k\to i} N_{j\to l}\bigr)^{\frac{1}{4}}.
\end{eqnarray}

A more generalised form of $\vert \textbf{w}_r\vert $ and $\lambda_r(\textbf{n})$ could be got:
\begin{eqnarray}
\vert \textbf{w}_r(\textbf{n})\vert  && = \sum_{i_1\to j_1, i_2\to j_2, \dots, i_r \to j_r} a_{i_1 j_1} M_{i_2\to i_1} N_{j_{r-1}\to j_r} \times\nonumber\\
&&\prod_{k = 1}^r n_{i_r} n_{j_r} \prod_{k = 1}^{r - 1} a_{i_k i_{k +1}} a_{j_k j_{k +1}} A_{i_k i_{k +1}} A_{j_k j_{k +1}},
\end{eqnarray}
and
\begin{eqnarray}
\lambda_r(\textbf{n}) && = \bigl(\frac{1}{2\vert E\vert } \sum_{i_1\to j_1, i_2\to j_2, \dots, i_r \to j_r} a_{i_1 j_1} M_{i_2\to i_1} N_{j_{r-1}\to j_r} \times\nonumber\\
&&\prod_{k = 1}^r n_{i_r} n_{j_r} \prod_{k = 1}^{r - 1} a_{i_k i_{k +1}} a_{j_k j_{k +1}} A_{i_k i_{k +1}} A_{j_k j_{k +1}}\bigr)^{\frac{1}{2r}}.
\end{eqnarray}

In our research we use the $r = 1$ case. The spreading influence of node $v_i$ is regarded as the contribution of all value of edge between $v_i$ and $v_j$ contained in $\lambda_1(\textbf{n})$, namely:
\begin{eqnarray}
SI_i && = \sum_{j\in N(i)} n_i n_j a_{ij} A_{ij} M_{i\to j} N_{i\to j}\nonumber\\
&& = \sum_{j\in N(i)} a_{ij}\sum_{k\in N(i), k\ne j}a_{ik} (d_j - 1).
\end{eqnarray}

\section{Spreading Simulation}\label{simulation}
To examine the validation of employment spreading influence, we apply an agent-based spreading simulation. We want to verify if the infection starts with high $empSI$ occupations will the it spread through the whole network faster.

The complete occupation networks in Dalton, GA and Los Angeles, CA are used to simulate the infection process. In the networks, nodes stand for occupations and edges are weighted by skill similarity. In each node, there are agents with the same number to employment of corresponding occupation. 

We select $0.2\%$ agents in Los Angeles and Dalton network scattered uniformly into top-20 occupations as initial infected agents. For each step, each infected agent (source agent) selected one agent from its neighbour occupations, which we call ``target agent''. If the target agent is not infected, the source agent infects the target with the probability of skill similarity; if the target agent is already infected, the source agent will do nothing. After each step, all the infected agent will become source agent and could start to infect others.

In different simulations, the initial infected agents are selected by highest $empSI$, highest employment and highest weighted degree. At the same time, a simulation with random selected initial infected agents is implemented to work as the baseline. For each simulation, 20 trials are implemented and average results are calculated. We present the relative infected numbers in each step against the baseline in the main paper.

The simulation process could be described as:
\begin{algorithm}[htbp]
	\caption{Infection Simulation Process}
	\begin{algorithmic}
	         \State {\bf Input:} Graph $G$, Seed Infection Set $S$
	         \State
	          \For {Each step}
	          \For{Any agent $i$ in $S$}
	          \State Selected one agent $j$ from the neighbours of $i$
	          \If{$j$ is not infected}
	          \State With probability $SkillSimilarity(i,j)$, infected $j$
	          \EndIf
	          \EndFor
	          \EndFor
	          \State Put all infected $j$ into $S$
	          \State
	          \State  {\bf Output:} Infected Sets $S$ in Each Step
	\end{algorithmic}
	\label{main_process}
\end{algorithm}

\begin{figure}[htbp]
	\centering
	\includegraphics[scale=0.32]{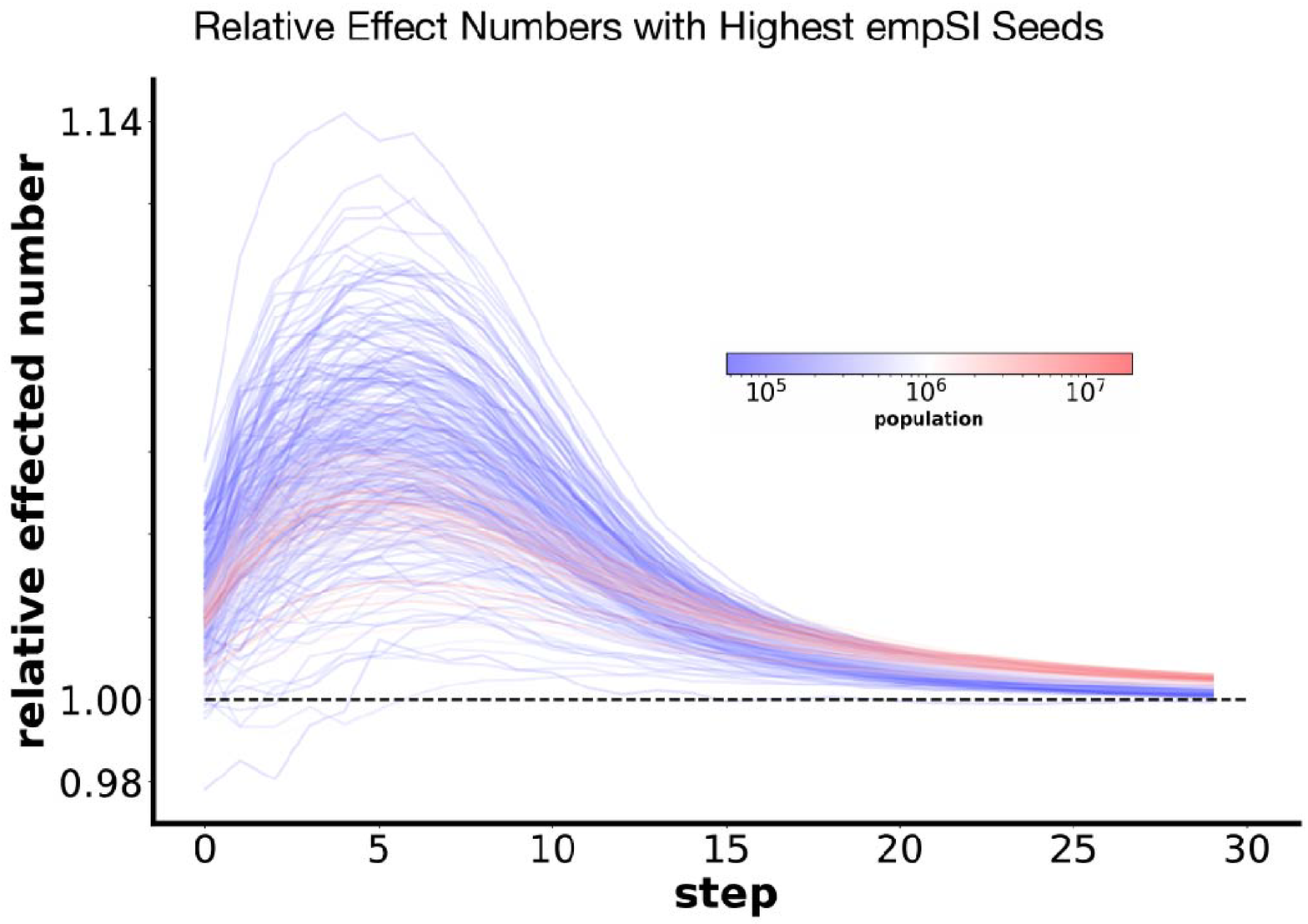}	
	\caption{Spreading simulation with high $empSI$ seeds against random results. Cities with different populations are marked by different colours, with blue stands for small cities and red large cities. Each curve is the average results of 20 simulation trials.}\label{empsi_all}
\end{figure}

Results of spreading simulation with highest $empSI$ jobs as seeds against random results in all cities are presented in Figure~\ref{empsi_all}. A subtle negative correlation between spreading speed and city scale could be observed: the spreading starting with high $empSI$ seeds tends to be slower than in smaller cities. We infer that this is due to the higher complexity and more diverse occupation structure in larger cities. Further exploration in the spreading dynamics geo-variation could be expected in the future.

\section{Correlation between Cognitive Scores and Spreading Influences}\label{cor_si_cs}
\begin{figure}[htbp]
	\centering
	\includegraphics[scale=0.28]{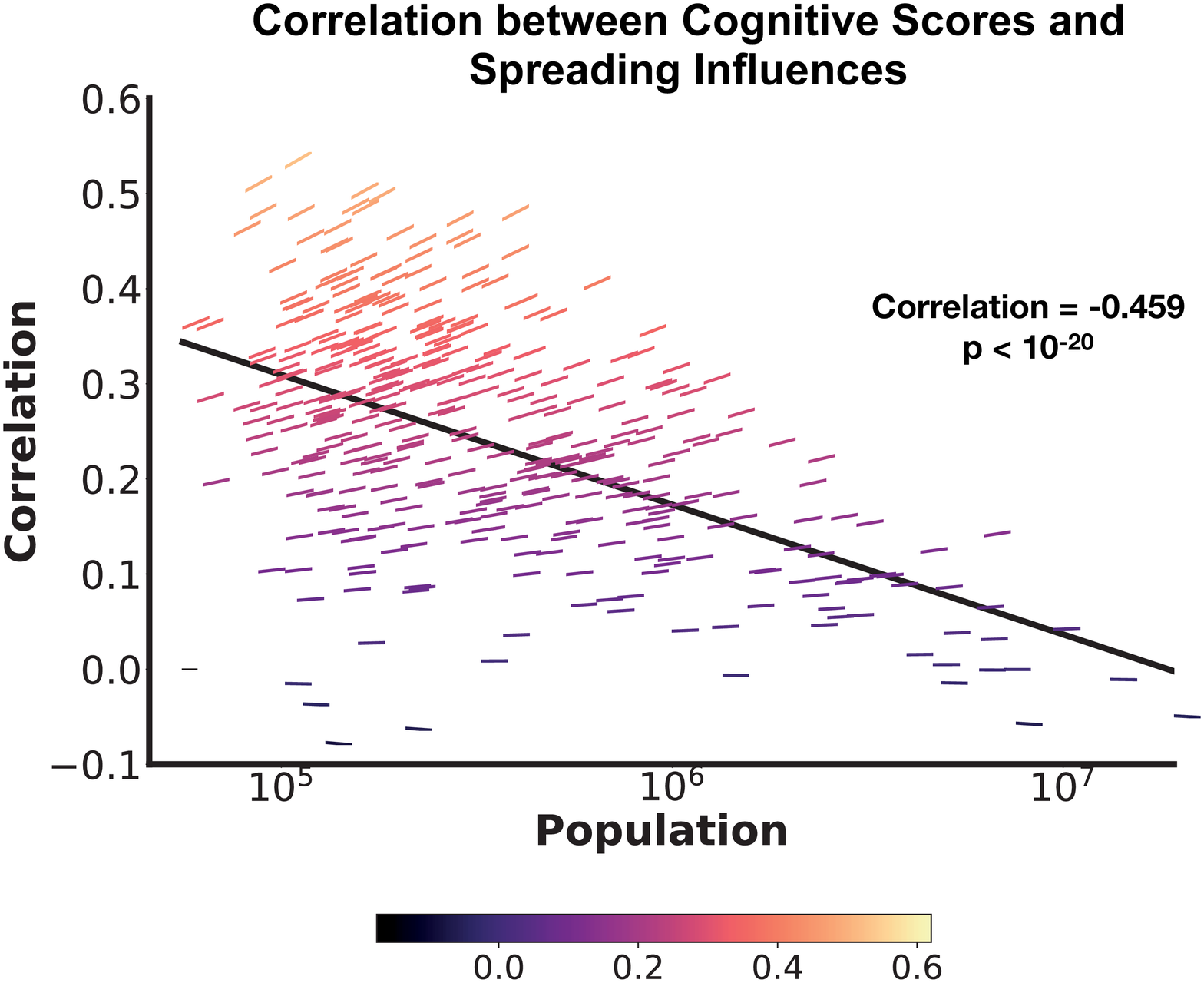}	
	\caption{Correlations between cognitive score values and $SI$ against populations in each city. The Pearson correlation coefficient is marked. Different colours and slopes stand for different correlation values.}\label{sicog}
\end{figure}

In the main text, we presented relationships between $empSI$ and cognitives in different cities. Here, a similar work is done for $SI$ and cognitive scores, see in Figure~\ref{sicog}. We find that there is a clear negative trend between the job-wise correlation in cognitive score and $SI$. This is despite smaller cities generally having fewer workers in cognitive jobs, suggesting that this is driven by the more numerous \textit{non-cognitive} jobs being relatively \textit{un-influential}. 

\section{Weighted Automation}\label{wauto}

\begin{figure}[htbp]
	\centering
	\includegraphics[scale=0.25]{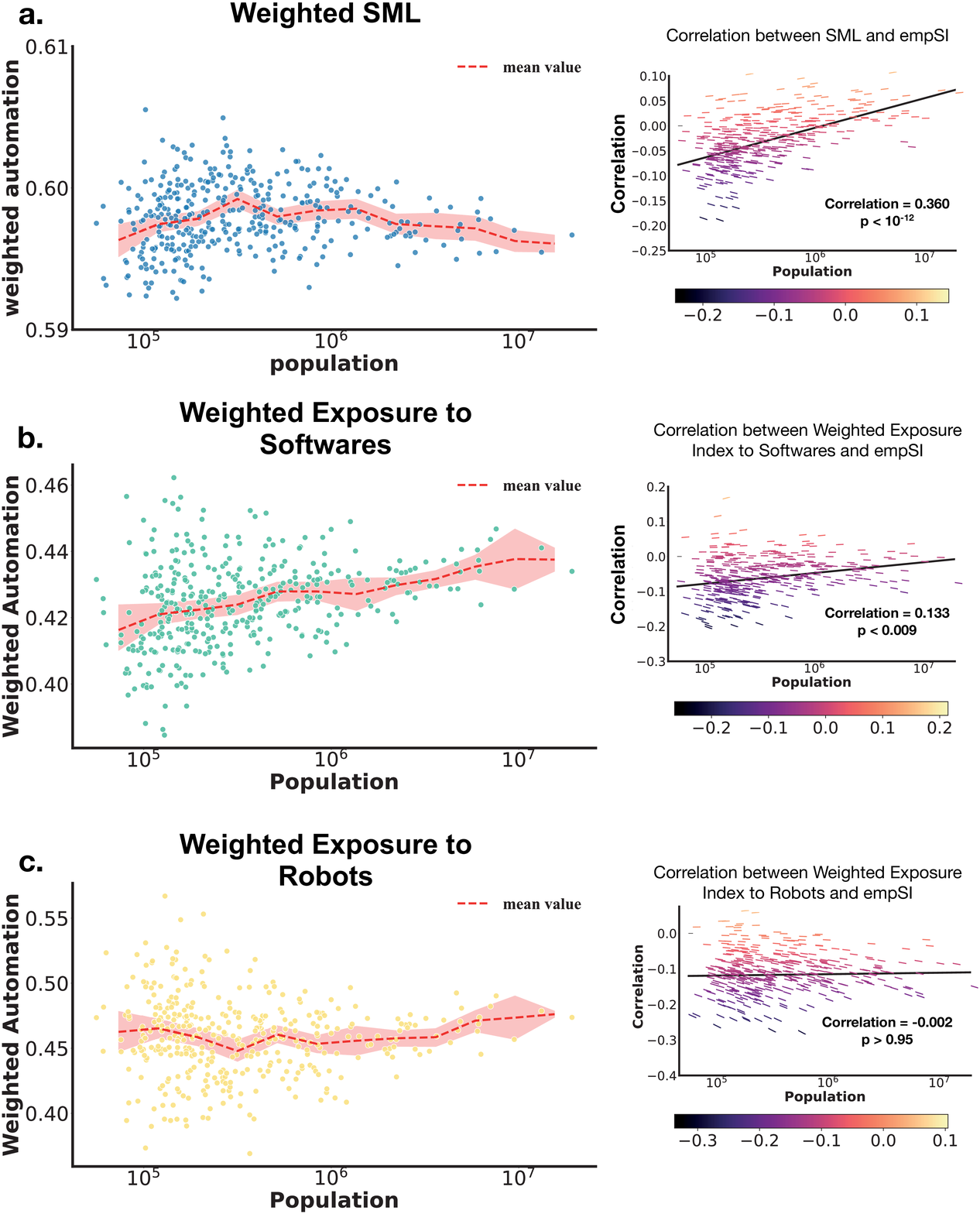}	
	\caption{\textbf{(a.)} Weighted suitability for machine learning (SML) from Brynjolfsson and Mitchell' research work with $empSI$. The correlation between SML and $empSI$ is drawn on the right side. \textbf{(b.)} Weighted exposure score to software patents. \textbf{(c.)} Weighted exposure score to robot patents }\label{siweightauto}
\end{figure}

In the main text, we calculate the weighted automation by $empSI$ with occupation data from Grey and Osborne's research work~\cite{frey2017future}. In their work, 70 occupations are selected to be hand-labelled their automatability by a group of machine learning researchers, which are used as the training set. Then nine variables from O*NET~\cite{onet} describing the occupations are selected to make the training with a Gaussian process classifier. In this way, the automation and computerisation probability of occupations from O*NET are estimated.

There are some other research works about automation probability estimation. Brynjolfsson and Mitchell designed a rubric with 23 distinct statements to evaluate the ``suitability for machine learning'' (SML)  of 2,069 direct work activities, where the occupations have been mapped to, to estimate the SML scores of occupations~\cite{brynjolfsson2018can}. Michael Webb calculate the overlaps between job task description texts and patent texts to measure the exposure of occupations to automation, in the perspectives of software, industrial robots and artificial intelligence. They both got meaningful data and conclusions~\cite{webb2019impact}.

Based on the automation data from these research, we also calculate their corresponding weighted automation by $empSI$ and results are presented in Figure \ref{siweightauto}. A similar conclusion could be seen from the SML data and larger cities perform higher resilience against automation. The curves from Michael Webb's data look opposite and larger cities present higher weighted values.

\section{Notes on RCA Values}\label{notes}
When we calculate the RCA values of $empSI$, there are some occupation with very high RCA values, mostly in the natural sources and production sectors. These high RCA values are mostly because the insufficient data, since these occupations do appear in other cities yet their employment numbers are not available. It is hard to decide whether the RCA values follow some certain distributions (like power law~\cite{barabasi1999emergence}, which will make the average meaningless). So in our research we remove some extreme high RCA values.




\end{appendices}

\bibliography{sn-bibliography}


\end{document}